\documentclass[prd,amsfonts,onecolumn,superscriptaddress,aps,nofootinbib,11pt]{revtex4-1}

\usepackage{booktabs}
\usepackage{amsmath, amsthm, amssymb, amsfonts, mathrsfs}
\usepackage{physics, braket, slashed, siunitx, bm}
\usepackage{bbold}
\usepackage{listings}
\usepackage{color}
\usepackage[usenames,dvipsnames]{xcolor}
\usepackage{graphicx}
\usepackage{float}
\usepackage{placeins}
\usepackage{booktabs}
\usepackage{hyperref}
\usepackage{soul}
\usepackage{ulem}
\def\ds{\partial\!\!\!/}
\def\As{A\!\!\!/}
\def\Bs{B\!\!\!/}

\def\bs{b\!\!\!/}
\def\ks{k\!\!\!/}
\def\ps{p\!\!\!/}

\hypersetup{
	unicode=false,          % non-Latin characters in Acrobat’s bookmarks
	pdftoolbar=true,        % show Acrobat’s toolbar?
	pdfmenubar=true,        % show Acrobat’s menu?
	pdffitwindow=false,     % window fit to page when opened
	pdfauthor={William},     % author
	colorlinks=true,       % false: boxed links; true: colored links
	linkcolor=blue,          % color of internal links
	citecolor=red,        % color of links to bibliography
	urlcolor=blue
}

\begin{document}

	\title{Axion-Photon Interaction from Nonminimal Dimension-5  Lorentz-Violating Operators }
	\author{A. J. G. Carvalho}
	\email{gomescarvalhoantoniojose@fisica.ufpb.br}
	\affiliation{ Departamento de Ciências Naturais,
		Universidade do Estado do Pará-UEPA, \\
		68502-100, Marabá, Pará, Brasil}
	
	\author{A. G. Dias}
	\email{alex.dias@ufabc.edu.br}
	\affiliation{Centro de Ciências Naturais e Humanas, Universidade Federal do ABC,~\\
		09210-580, Santo André, São Paulo, Brasil}
	\author{A. F. Ferrari}
	\email{alysson.ferrari@ufabc.edu.br}
	
	\affiliation{Centro de Ciências Naturais e Humanas, Universidade Federal do ABC,~\\
		09210-580, Santo André, São Paulo, Brasil}
	\author{T. Mariz}
	\email{tmariz@fis.ufal.br}
	
	\affiliation{Instituto de Física, Universidade Federal de Alagoas,~\\
		57072-900, Maceió, Alagoas, Brasil}
	\author{J. R. Nascimento}
	\email{jroberto@fisica.ufpb.br}
	
	\affiliation{Departamento de Física, Universidade Federal da Paraíba,~\\
		Caixa Postal 5008, 58051-970, João Pessoa, Paraíba, Brasil}
	\author{A. Yu. Petrov}
	\email{petrov@fisica.ufpb.br}
	
	\affiliation{Departamento de Física, Universidade Federal da Paraíba,~\\
		Caixa Postal 5008, 58051-970, João Pessoa, Paraíba, Brasil}
	\date{\today}
	\begin{abstract}
		In this paper, we discuss various possible schemes for the perturbative
		generation of the axion-photon interaction term in different Lorentz-breaking
		extensions of QED, involving operators with mass dimensions up to
		5. We demonstrate explicitly that there are only a few schemes allowing one
		to generate a finite axion-photon interaction term from one-loop radiative corrections,
		and within all these schemes, the generated term turns out to be ambiguous. 
		Also, through the current experimental limits on the axion-photon 
		coupling, we extract some present constraints on the Lorentz violation parameters. 
	\end{abstract}
	\maketitle
	
	\section{Introduction}
	
	Axions and axion-like particles (ALPs) are light pseudoscalar particles predicted in several models aiming to solve problems left open by the Standard Model. Generically, they are pseudo-Nambu-Goldstone bosons of a spontaneously broken approximate global symmetry, with their mass arising from supposedly suppressed violating symmetry operators. Axions are associated with the U(1)$_{PQ}$ Peccei-Quinn symmetry, which allows to obtain a solution to the strong CP problem~\cite{Peccei:1977hh,Weinberg:1977ma,Wilczek:1977pj}, i.e. the absence of interactions violating charge conjugation and parity symmetries in the strong interactions. As it happens, the axion mass and couplings are related, being all inversely proportional to the energy scale in which the Peccei-Quinn symmetry breaks down. In the invisible axion models~\cite{Kim:1979if,Shifman:1979if,Dine:1981rt,Zhitnitsky:1980tq}, this scale can be high enough to imply a low mass and very weak interactions for the axion, which may also figure as a dark matter candidate~\cite{Abbott:1982af,Preskill:1982cy,Dine:1982ah}. The name ALP is, in fact, a generic designation for particles with similar features to the axion, except that their mass and couplings are unrelated. The majoron, which is associated with the lepton number symmetry~\cite{Chikashige:1980ui,Gelmini:1980re}, is one example of ALP. Both axions and ALPs with faint interactions are general predictions of string theory, since they naturally arise within the context of string compactifications, as reviewed in ~\cite{Ringwald:2012cu}. It is also worthwhile to remark that the electrodynamics with
	the axion term is also studied within the topological insulator context
	\cite{Martin-Ruiz:2015skg,Martin-Ruiz:2016qms}. Once the difference between axions and ALPs is not so relevant to this work, we will refer to these particles simply as ``axions'' without making any distinction among them in what follows.  
	
	The axion-photon interaction is of central importance for investigating observable axion effects. It has been tested in laboratory searches, astrophysics, and cosmology, as we can  see from the compilation in  Refs.~\cite{Sikivie:2020zpn,DiLuzio:2020wdo,ParticleDataGroup:2022pth}. One way of obtaining the axion-photon interaction term is to start
	with an axion field $\phi$ coupled to a charged fermion $\psi$, through the parity-violating interaction term   $i\phi\overline{\psi}\gamma_{5}\psi$.
	Following the general methodology of the low-energy effective theory,
	integration of the fermion field leads to an effective Lagrangian
	describing the pseudoscalars interacting with photons of the form
	\begin{equation}
		\mathcal{L}_{{eff}}=-\frac{1}{4}F_{\mu\nu}F^{\mu\nu}+\frac{1}{2}\partial_{\mu}\phi\partial^{\mu}\phi-\frac{m_{\phi}}{2}\phi^{2}-\frac{g_{\phi\gamma}}{4}\phi F_{\mu\nu}\widetilde{F}^{\mu\nu}\,,\label{eq:leff}
	\end{equation}
	where $g_{\phi\gamma}$ is the axion-photon coupling constant, which has dimension of  inverse of mass; $F_{\mu\nu}$ is the electromagnetic field strength, with
	$\tilde{F}^{\mu\nu}=\frac{1}{2}\varepsilon^{\mu\nu\alpha\beta}F_{\alpha\beta}$
	its dual. The last term in Eq. (\ref{eq:leff})  is the most common form for
	the axion-photon interaction term.

	It was shown in ~\cite{Borges:2013eda} that the axion-photon term can also be generated as radiative corrections within a particular nonminimal -- including interactions of  higher orders in derivatives -- Lorentz violating (LV) extension of QED. In such a context, the generated axion term is finite but ambiguous, i.e. its value depends on the computation scheme, explicitly, on the regularization used within the calculations. 
	A similar ambiguity has been intensively discussed in the literature
	regarding the generation of the Carroll-Field-Jackiw (CFJ) term \cite{Carroll:1989vb},
	which is an LV generalization of the Chern-Simons term to four space-time
	dimensions (see also~\citep{Altschul:2019eip}, and references therein).
	This result establishes an interesting relationship between two very active
	fields of research, namely, the search for axions, and
	the systematic investigation of the possibility of Lorentz Violation
	that has been guided by the formulation of the Standard Model Extension
	(SME)\,\citep{Colladay:1998fq}. This points out the possibility of constraining LV interactions via the effective axion-photon interaction. 
	
	We are interested in studying more generally the mechanism
	of radiative generation of  the axion-photon interaction term, using as a guide the generalization
	of the QED sector of the SME to include nonminimal interaction terms, as described
	in\,\cite{Kostelecky:2018yfa}. We want to find out if the mechanism
	uncovered in ~\cite{Borges:2013eda} is unique and, if other instances
	of radiative generation of a finite axion-photon interaction term exist, whether they
	suffer of the same kind of ambiguity, as it was found there. We will
	show that the answer to both questions is positive: there are different ways 
	to generate a finite axion-photon interaction
	term, but all of them suffer from the same kind of ambiguity. To keep
	our search reasonably contained, we will restrict ourselves to LV
	operators of mass dimension up to five, i.e., the dimension of Lorentz-breaking
	parameters is no less than $-1$.
	
	The paper is structured as follows. In section~\ref{IS}, we
	formulate the initial statements of our problem and outline the calculations
	we will make to answer the questions we pose. In section~\ref{OLC},
	we identify the situations in which the generation of the axion-photon interaction term
	occurs, and perform one-loop calculations to verify its appearance. Some constraints on the Lorentz violation parameters entering in the axion-photon coupling are present in section \ref{lvplimits}. Finally, in section~\ref{SUM}, we formulate our conclusions.
	
	\section{Initial Statements\label{IS}}

	Our aim consists in finding the possibilities to generate the axion-photon
	interaction term in various Lorentz-breaking extensions of QED. By
	definition, the axion term we are looking for is given by 
	\begin{equation}
		{\cal L}_{axion}^{\left(1\right)}=-\frac{g_{\phi\gamma}}{8}\phi\epsilon^{\mu\nu\lambda\rho}F_{\mu\nu}F_{\lambda\rho}=g_{\phi\gamma}\phi\vec{E}\cdot\vec{B}\thinspace,\label{axion}
	\end{equation}
	which is explicitly Lorentz invariant (LI), with the coupling $g_{\phi\gamma}$ depending on LV parameters. So, first we have to explain
	in what sense this term can be generated from a Lorentz violating
	model. One possibility, that was uncovered in\,\cite{Borges:2013eda},
	is a scheme involving two LV vectors that are contracted, leading
	to a LI axion term. Specifically, it has been shown that radiative
	corrections involving the fermion-photon nonminimal interaction $d^{\nu}F_{\mu\nu}\overline{\psi}\gamma^{\mu}\psi$
	and the LV Yukawa coupling $b_{\mu}\phi\bar{\psi}\gamma_{5}\gamma^{\mu}\psi$
	generates the axion-photon interaction in the form 
	\begin{equation}
		{\cal L}_{axion}^{\left(1\right)}=C\,e\,\phi\epsilon^{\mu\alpha\beta\rho}d_{\rho}b^{\kappa}F_{\alpha\beta}F_{\kappa\mu}=2C\,e(b\cdot d)\thinspace\phi\thinspace(\vec{E}\cdot\vec{B})\thinspace,\label{axion1}
	\end{equation} 
 where $e$ is the electromagnetic coupling and $C$ the mentioned ambiguous constant whose value depends on the regularization scheme.
	
	Another possibility is to notice that the CFJ term \cite{Carroll:1989vb}
	can be related to the axion by assuming the axion field $\phi$
	to be slowly varying, so that $b_{\mu}=\partial_{\mu}\phi$ can
	be considered as a small and constant LV vector. This relation is
	given by the equality 
	\begin{equation}
		-\frac{1}{4}\phi\tilde{F}F=\frac{1}{2}\epsilon^{\mu\nu\lambda\rho}b_{\mu}A_{\nu}F_{\lambda\rho}\thinspace,
	\end{equation}
	which holds up to surface terms. Now, it is a very well known fact
	(see, e.g., \cite{Colladay:1998fq}) that the CFJ term (the right-hand
	side of this last equation) can be generated by the $b_\mu$ term in the
	minimal LV extension of QED. That is, Eq.\,\eqref{axion} %\textcolor{blue}{[ou Eq.\,\eqref{eq:leff}??]} 
	could naturally
	emerge from quantum corrections starting with 
	\begin{equation}
		{\cal L}=\bar{\psi}(i\ds-m-\bs\gamma_{5}-e\As)\psi-\frac{1}{4}F_{\mu\nu}F^{\mu\nu}.\label{minex}
	\end{equation}
	Interesting as it is, we may argue that this scheme is not the one
	we are interested in. First, it demands the axion field $\phi$
	to be considered as a (fixed) slowly varying background (it worth mentioning that the relation between spacetime-varying constants and Lorentz symmetry breaking has been originally claimed in \cite{Kostelecky:2002ca}). Actually,
	gauge invariance of the CFJ term demands $\phi=k_{\lambda}x^{\lambda}$ so that $k_{\lambda}\propto b_{\lambda}$,
	with $k_{\mu}$ being a (very small) constant. On the other hand,
	if we want to make connection with the experimental searches
	for axions/ALPs, it is more natural to consider $\phi$ as
	an arbitrary, dynamic field. Second, there are arguments that the
	ambiguous CFJ term generated via this mechanism should vanish, in
	order to preserve gauge invariance\,\cite{Altschul:2019eip}. For
	these reasons, we will refer to this as the ``trivial scheme'',
	which will not be our main concern.
	
	Therefore we will concentrate on looking for possible schemes of generation
	of terms similar to (\ref{axion1}), disregarding the trivial scheme
	outlined above. This effectively means that we will study contributions
	involving one constant vector and one constant axial vector, that will eventually appear in contracted form in order to reproduce
	the LI axion term given in (\ref{axion}). Since the axion is pseudoscalar,
	the couplings involved in the scheme should contain either $\gamma_{5}$
	or $\epsilon^{\mu\nu\lambda\rho}$. This will be an important guide
	in our considerations.
	
	Following the methodology of constructing Lorentz-breaking extensions
	of QED \cite{Colladay:1998fq,Kostelecky:2018yfa}, we start by defining
	the following generic LV extension of QED, 
	\begin{equation}
 \label{genextension}
		{\cal L}=\bar{\psi}\left(i\ds-m-e\As+\frac{1}{\Lambda^{n-4}}\hat{T}^{(n)}\right)\psi-\frac{1}{4}F_{\mu\nu}F^{\mu\nu}\thinspace,
	\end{equation}
	where $\hat{T}^{(n)}$ is a dimension-$n$ LV operator. To arrive at the axion-photon interaction term in Eq. (\ref{axion1}), after the loop calculations one can multiply the resulting $A_{\mu}$-dependent term by $\phi$, which is equivalent to replacement of $T^{(n)}$ by $\phi T^{(n)}$. Actually this corresponds to extracting the leading order term in an expansion in powers of the momenta associated to the $\phi$ field, as explained in detail in \,\citep{Borges:2013eda}.
	In what follows, we assume this substitution to be taken from the very beginning.
	It is clear that the axion term arises if the operator $\hat{T}^{(n)}$ involves
	either the Levi-Civita symbol or the $\gamma_{5}$ matrix. The operators
	we will consider are extracted from the general classification presented
	in \cite{Kostelecky:2018yfa}. Finally, we restrict our consideration
	for Lorentz-breaking operators of dimension 5, i.e., $n=5$. Higher
	dimensions will not be considered for simplicity, and the restriction
	to LV operators of mass dimension three or four corresponds to the
	minimal LV extension of QED\,\cite{Colladay:1998fq,Kostelecky:2001jc},
	\begin{equation}
		{\cal L}_{mQED}=\frac{i}{2}\overline{\psi}\Gamma^{\nu}\overleftrightarrow{D}_{\mu}\psi-\overline{\psi}M\psi+{\cal L}_{photon}\thinspace,
	\end{equation}
	where ${\cal L}_{photon}$ is the minimal LV Maxwell Lagrangian, 
	\begin{align}
		\Gamma^{\nu} & \equiv\gamma^{\nu}+c^{\mu\nu}\gamma_{\mu}+d^{\mu\nu}\gamma_{5}\gamma_{\mu}+e^{\nu}+if^{\nu}\gamma_{5}+\frac{1}{2}g^{\lambda\mu\nu}\sigma_{\lambda\mu}\thinspace,\\
		M & \equiv m+a_{\mu}\gamma^{\mu}+b_{\mu}\gamma_{5}\gamma^{\mu}+\frac{1}{2}H_{\mu\nu}\sigma^{\mu\nu}\thinspace,
	\end{align}
	and $\sigma_{\mu\nu}$ is the commutator of two $\gamma$ matrices.  Here $a_{\mu}$, $b_{\mu}$, $c_{\mu\nu}$, $d_{\mu\nu}$, $e_{\mu}$, $f_{\mu}$, $g_{\lambda\mu\rho}$, $H_{\mu\nu}$ are constant tensors of corresponding ranks, introducing the Lorentz symmetry breaking. All these parameters have non-negative mass dimensions, hence they do not jeopardize the renormalizability of the theory.
	%\textcolor{blue}{Explicar os coeficientes.}
	
	It is easy to see that in this minimal case, no axion term can be
	generated except by the trivial scheme. Indeed the only operators
	involving $\gamma_{5}$ are those proportional to $b_{\mu}$,
	$d_{\mu\nu}$ and $f_{\mu}$, and the absence of contributions proportional
	to the first order in $d_{\mu\nu}$ and $f_{\mu}$ is checked
	straightforwardly. We note that this conclusion matches the results
	found in \cite{Kostelecky:2001jc}, while also being consistent with
	\cite{BaetaScarpelli:2015ykr}. As for the possible second-order terms
	of the form (\ref{axion1}), we note that since they require one vector
	and one pseudovector, they must be proportional to products of $b_{\mu}$,
	$f_{\mu}$ or $h_{\mu}=\epsilon_{\mu\nu\lambda\rho}g^{\nu\lambda\rho}$  (here, for the sake of simplicity, we choose the $g^{\nu\lambda\rho}$ to be completely antisymmetric, and such a choice has been used earlier in \cite{BaetaScarpelli:2018vsm})
	with $e_{\mu}$. Such contributions can be easily shown to vanish. Indeed, the vector $e_{\mu}$ enters the action only within the contraction with $A_{\mu}$ or $\partial_{\mu}$. In the first case, the only possible second-order terms involving the Levi-Civita symbol and second orders in gauge fields and derivatives should look like $(e\cdot A)\epsilon^{\mu\nu\lambda\rho}q_{\mu}\partial_{\nu}\partial_{\lambda}A_{\rho}$, where $q_{\mu}$ is for any of our pseudovectors, i.e., $b_{\mu}$, $f_{\mu}$ or $h_{\mu}$. Such a term is evidently equal to zero. In the second case, vanishing of corresponding contributions is shown straightforwardly.
	
	\section{One-loop calculations\label{OLC}}
	
	So, let us begin with our calculations. Actually, we must consider
	two situations: (i) corrections of the first order in LV parameters
	and (ii) corrections of the second order in LV parameters.
	
	\subsection{First-order contributions}
	
\begin{figure}[h]
		\centering{}\includegraphics[scale=0.9]{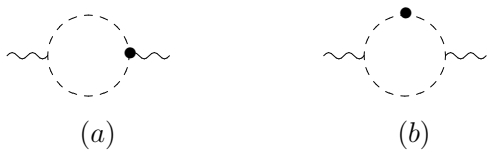} \caption{General form of first order Lorentz-breaking contributions.}
		\label{Fig1}
	\end{figure}

	Let us start with the first scenario. It is clear that the axion-photon terms of the first order in LV parameters must have the form $\phi\epsilon^{\mu\alpha\beta\rho}t_{\rho}^{\phantom{\rho}\kappa}F_{\alpha\beta}F_{\kappa\mu}$, which reduces to the standard expression (\ref{axion1}) in the case $t_{\rho\kappa}=d_{\rho}b_{\kappa}$. Effectively, to obtain the lower  contributions, those ones of the first order in LV parameters, we have two possibilities: first, we replace one of the vertices in the usual contribution to the two-point function of the gauge field by a nonminimal one, or, second, we consider a nonminimal modification of the kinetic term of the spinor field, which implies in a LV insertion in the spinor propagator. These two possibilities can be formally represented by the Feynman diagrams given in Fig.~\ref{Fig1}, where the black dot is for a nonminimal LV insertion.
	
	\vspace*{3mm}
	
	\vspace*{3mm}
	
	Now, let us list the possible dimension-5 operators. Following\,\citet{Kostelecky:2018yfa},
	we have either operators proportional to $iD_{(\alpha}iD_{\beta)}$
	(where as usual $D_{\alpha}=\partial_{\alpha}+ieA_{\alpha}$) or operators
	proportional to $F_{\alpha\beta}$. Let us start with the second case.
	It is clear that if we consider operators involving $F_{\alpha\beta}$,  the propagator of $\psi$ is not modified, but a new vertex proportional to $F_{\alpha\beta}$ is present. Therefore, the corresponding diagrams can have only the form (a).  Moreover, as we consider the
	first-order contributions, we assume that we have no other LV insertions.
	Also, it is evident that the terms proportional to odd-rank constant
	tensors can be disregarded.
	
	Immediately we see that the operator $H_{F}^{(5)\mu\nu\alpha\beta}\bar{\psi}\sigma_{\mu\nu}\psi F_{\alpha\beta}$
	cannot generate the axion-photon term (except for the trivial case $H_{F}^{(5)\mu\nu\alpha\beta}\propto\epsilon^{\mu\nu\alpha\beta}$,
	which actually is LI and therefore outside of our  interest).
	The operator proportional to $-\frac{1}{2}m_{F}^{(5)\alpha\beta}\bar{\psi}F_{\alpha\beta}\psi$,   
	in the case
	of the absence of other insertions, is ruled out as well (i.e., the
	scalar, vector, and tensor couplings cannot yield axion-photon contributions
	since they do not allow for the appearance of the Levi-Civita symbol,
	unlike axial couplings).
	
	So, it remains to consider axial couplings. We start with dimension-5
	vertices proportional to $F_{\alpha\beta}$. The only even-rank axial
	coupling is represented by the term $-\frac{i}{2}m_{5F}^{(5)\alpha\beta}\bar{\psi}\gamma_{5}F_{\alpha\beta}\psi$. The graph (a), after expansion of the corresponding propagator
	in the external momentum, is proportional to 
	\begin{eqnarray}
		\Sigma_{5F}(p) & = & {\rm tr}\int\frac{d^{4}k}{(2\pi)^{4}}\gamma_{\lambda}(\ks+m)\gamma_{5}(\ks+m)\ps(\ks+m)\frac{1}{(k^{2}-m^{2})^{3}}A^{\lambda}(-p)m_{5F}^{(5)\alpha\beta}F_{\alpha\beta}(p),
	\end{eqnarray}
	which is zero, either by vanishing of the integral or by vanishing
	of the trace, for any form of $m_{5F}^{(5)\alpha\beta}$. 
	
	As for the terms proportional to $\bar{\psi}iD_{(\alpha}iD_{\beta)}\psi$,
	we can apply the same arguments as above to prove that such terms
	could yield the axion-photon form (\ref{axion1}) only if they are proportional
	to the second-rank constant pseudotensor. From the classification presented in \cite{Kostelecky:2018yfa},
	the only candidate is $im_{5}^{(5)\alpha\beta}\thinspace\bar{\psi}\gamma_{5}iD_{(\alpha}iD_{\beta)}\psi$.
	However, the pseudotensor $m_{5}^{(5)\alpha\beta}$ is symmetric.
	It is easy to show that all scalars involving the Levi-Civita symbol,
	two field strengths, and a symmetric $m_{5}^{(5)\alpha\beta}$, identically vanish. Therefore, there is no axion-photon contribution proportional to $m_{5}^{(5)\alpha\beta}$.

	As a result, we conclude that there is no generation of the axion
	term of the first order in LV parameters.
	
	\subsection{Second-order contributions}
	
	Let us consider the possibility of the generation of an axion term of
	second-order in LV coefficients. Within our study, we restrict ourselves
	to those ones generated by two vertices, with one of them proportional
	to a constant LV vector and another one to a constant LV pseudovector,
	to match the form (\ref{axion1}). Essentially, we are looking for
	possible generalizations of the mechanism outlined in\,\cite{Borges:2013eda}.
	This also means that we can consider terms from \cite{Kostelecky:2018yfa}
	proportional to any odd-rank tensors as well since they, in certain
	cases, can be treated as contractions of constant (pseudo)vectors
	either with the Minkowski metric or with the Levi-Civita symbol. 
	Explicitly, we focus on the terms: $-\frac{1}{2}a^{(5)\mu\alpha\beta}_F\bar{\psi}\gamma_{\mu}F_{\alpha\beta}\psi$, $-\frac{1}{2}b^{(5)\mu\alpha\beta}_F\bar{\psi}\gamma_5\gamma_{\mu}F_{\alpha\beta}\psi$,  $-\frac{1}{2}a^{(5)\mu\alpha\beta}\bar{\psi}\gamma_{\mu}iD_{(\alpha}iD_{\beta)}\psi$, and $-\frac{1}{2}b^{(5)\mu\alpha\beta}\bar{\psi}\gamma_5\gamma_{\mu}iD_{(\alpha}iD_{\beta)}\psi$.

	First, there is a possibility for a non-zero contribution involving
	$a_{F}^{(5)\mu\alpha\beta}$. We consider the case $a_{F}^{(5)\mu\alpha\beta}=\epsilon^{\mu\alpha\beta\gamma}b_{\gamma}$,
	corresponding to the known dimension-5 operator $\bar{\psi}\epsilon^{\mu\nu\lambda\rho}\gamma_{\mu}b_{\nu}F_{\lambda\rho}\psi$,
	which has been used in \cite{Gomes:2009ch} to generate a finite aether
	term. In this case, $b_{\nu}$ is a pseudovector. The contribution
	of the first order in $b_{\nu}$ has been evaluated in \cite{Mariz:2021cik},
	where it was shown to yield a finite and ambiguous result for the
	CFJ term (it should be noted that, since the dimension-5 vertices, including the Lorentz-breaking case, are non-renormalizable, their presence can in principle imply in arising divergent contributions in other sectors of the effective action, both purely gauge ones and those ones including the spinor fields, such contributions will be studied elsewhere). However, as noted above, arising of the CFJ term corresponds to the
	trivial scheme commented in Section\,\ref{IS}. So, now we proceed
	with the second-order contributions involving this pseudovector.
	
	For the next step, we consider the following LV extension of the QED involving two such
	nonminimal vertices: 
	\begin{eqnarray}
 \label{twononmin}
		S=\int d^{4}x\thinspace\bar{\psi}(i\ds-m-e\As+a^{\alpha}F_{\alpha\beta}\gamma^{\beta}+\epsilon^{\alpha\beta\gamma\delta}b_{\alpha}F_{\beta\gamma}\gamma_{\delta})\psi.
	\end{eqnarray}
Here $a^{\mu}$ and $b^{\nu}$ are the usual (polar) and axial vectors,
	respectively.
	
	First, we obtain the contribution involving one minimal vertex and
	one magnetic-like vertex, that is, $\bar{\psi}\epsilon^{\mu\nu\rho\sigma}b_{\mu}F_{\nu\rho}\gamma_{\sigma}\psi$, which is of first order in Lorentz-breaking parameter $b_{\mu}$. The corresponding Feynman diagram has been evaluated in \cite{Mariz:2021cik},
	where it was demonstrated that in this case the finite and ambiguous
	CFJ term arises. 
	This is another instance of the trivial mechanism
	we are not interested in.
	
\begin{figure}[h]
		\centering{}\includegraphics[scale=0.9]{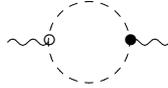} \caption{Contributions involving two different nonminimal LV vertices.}
		\label{Fig2}
	\end{figure}

	We can also obtain the axion-photon term involving two nonminimal vertices
	and contributing to the second-order in Lorentz-breaking parameters,
	represented in Fig.~\ref{Fig2}.
	
 \noindent 
 The contribution of this diagram looks like 
	\begin{eqnarray}
		\Sigma_{ab}(p) & = & -\epsilon^{\alpha\beta\lambda\rho}b_{\alpha}F_{\beta\lambda}(-p)a_{\alpha'}F^{\alpha'\rho'}(p)\int\frac{d^{4}k}{(2\pi)^{4}}\frac{1}{(k^{2}-m^{2})^{2}}{\rm tr}\big[m^{2}\gamma_{\rho}\gamma_{\rho'}+\nonumber \\
		& + & k^{\mu}k^{\nu}\gamma_{\mu}\gamma_{\rho}\gamma_{\nu}\gamma_{\rho'}\big].
	\end{eqnarray}
	After doing all contractions and integration, returning to Minkowski
	space, we get 
	\begin{eqnarray}
\Sigma_{ab}=-2m^{2}C_{1}a^{\alpha}F_{\alpha\gamma}b_{\beta}\tilde{F}^{\beta\gamma},\label{finamb}
	\end{eqnarray}
	with $C_{1}$ being the ambiguous finite constant defined in \cite{Gomes:2009ch}. Its explicit value can be read off from the expression:
		\begin{eqnarray}
			m^2C_1\eta_{\gamma\delta}=\int\frac{d^4k}{(2\pi)^4}\frac{1}{(k^2-m^2)^2}\left[m^2\eta_{\gamma\delta}+k^{\alpha}k^{\beta}(\eta_{\alpha\gamma}\eta_{\beta\delta}+\eta_{\beta\gamma}\eta_{\alpha\delta}-\eta_{\alpha\beta}\eta_{\gamma\delta})\right].
		\end{eqnarray}
		Within various regularization schemes, this constant is known to take different values, e.g., if we apply the symmetrization 
		$k^{\mu}k^{\nu}\to\frac{1}{4}\eta^{\mu\nu}k^2$ and afterwards promote the integral to $4+\epsilon$ dimensions, the result will be $C_1=\frac{1}{8\pi^2}$ \cite{Gomes:2009ch}. If we apply the symmetrization $k^{\mu}k^{\nu}\to\frac{1}{4+\epsilon}\eta^{\mu\nu}k^2$ and integrate in $4+\epsilon$ dimensions, the result will be $C_1=0$  \cite{BaetaScarpelli:2013rmt}.
	That is, proceeding with the substitution $a^{\alpha}F_{\alpha\beta}\gamma^{\beta}\rightarrow a^{\alpha}\phi F_{\alpha\beta}\gamma^{\beta}$, 
  as mentioned above, we succeeded in obtaining the axion-photon interaction of
	the form (\ref{axion1}), which is finite and ambiguous, i.e., the result for the loop integral depends on the regularization scheme.

	Although this case is very interesting since it allows us to demonstrate
	the generation of a finite and ambiguous axion-photon term, we note
	that any nonminimal vertex yields a very small contribution, since
	it is a general expectation that the higher the mass dimension of
	an operator, more suppressed it should be. However, if the fermion $\psi$ would be heavy enough, in the sense that $m^{2}C_{1}a^{\alpha}b_{\beta}$ may be comparable to the typical axion-photon coupling in models without LV, then Eq.~(\ref{finamb}) could represent a relevant contribution. Anyway, it is also interesting
	to generate axion-photon contributions involving one nonminimal LV insertion
	and another minimal one.  For the sake of concreteness, we will consider
	the cases where the minimal LV term is the most studied one, i.e., $\bar{\psi}\bs\gamma_{5}\psi$. That means, we will consider two additional ways to generate
	the axion term: first, the insertion of $\bs\gamma_{5}$ into a spinor
	propagator, together with one vertex proportional to $F_{\mu\nu}$,
	second, insertion of the same $\bs\gamma_{5}$ into a spinor propagator
	with the presence of the vertex proportional to $iD_{(\alpha}iD_{\beta)}$.
	
	\subsubsection{Second-order contributions with $\bs\gamma_{5}$ and $F_{\mu\nu}$}
	
	Let us start with contributions involving $F_{\mu\nu}$. Straightforward
	calculations show that in the case of inserting $\bs\gamma_{5}$ into
	a spinor propagator, the contributions proportional to $m_{F}^{(5)\alpha\beta}$
	and $m_{5F}^{(5)\alpha\beta}$ 
	(as well as to other even-rank constant
	tensors) vanish. Actually, the first promising dimension-5 operator
	cited in \cite{Kostelecky:2018yfa} in this case is $\bar{\psi}b_{F}^{(5)\mu\alpha\beta}\gamma_{5}\gamma_{\mu}F_{\alpha\beta}\psi$.
	The contribution involving this operator, together with the $\bs\gamma_{5}$
	insertion, can yield the axion-photon interaction only if $b_{F}^{(5)\mu\alpha\beta}=\epsilon^{\mu\alpha\beta\gamma}n_{\gamma}$,
	with $n_{\gamma}$ being a vector. In this case one can write 
	\begin{eqnarray}
		\Sigma_{bn}(p) & = & \epsilon^{\mu\rho\sigma\gamma}n_{\gamma}F_{\rho\sigma}(-p)A^{\lambda}(p)\nonumber \\
		& \times & {\rm tr}\int\frac{d^{4}k}{(2\pi)^{4}}\Big[\gamma_{5}\gamma_{\mu}(\ks+m)\bs\gamma_{5}(\ks+m)\gamma_{\lambda}(\ks+m)\ps(\ks+m)\nonumber \\
		& + & \gamma_{5}\gamma_{\mu}(\ks+m)\gamma_{\lambda}(\ks+m)\bs\gamma_{5}(\ks+m)\ps(\ks+m)\nonumber \\
		& + & \gamma_{5}\gamma_{\mu}(\ks+m)\gamma_{\lambda}(\ks+m)\ps(\ks+m)\bs\gamma_{5}(\ks+m)\Big]\frac{1}{(k^{2}-m^{2})^{4}}\thinspace.\label{eq:contrib}
	\end{eqnarray}
	This expression superficially logarithmically diverges. Next, we use
	the trace identity 
	\begin{equation}
		{\rm tr}(\gamma_{\alpha}\gamma_{\beta}\gamma_{\gamma}\gamma_{\delta})=4(\eta_{\alpha\beta}\eta_{\gamma\delta}-\eta_{\alpha\gamma}\eta_{\beta\delta}+\eta_{\alpha\delta}\eta_{\beta\gamma}).
	\end{equation}
	Calculating traces and symmetrizing the integrals through the four-dimensional
	replacements $k_{\alpha}k_{\beta}\to\frac{1}{4}\eta_{\alpha\beta}k^{2}$
	and $k_{\alpha}k_{\beta}k_{\gamma}k_{\delta}\to\frac{1}{24}(\eta_{\alpha\beta}\eta_{\gamma\delta}+\eta_{\alpha\gamma}\eta_{\beta\delta}+\eta_{\alpha\delta}\eta_{\beta\gamma})$,
	we find that the divergent part of this contribution, already before
	the integration, is zero. Such vanishing of the divergent part can indicate that the result
	is ambiguous,  which is very typical in LV theories (see, e.g., \cite{Gomes:2007rv}).
	At the same time, the superficially finite part of this contribution differs from zero, being equal to
	\begin{equation}
		\Sigma_{bn}=-\frac{1}{48\pi^{2}}b^{\mu}F_{\mu\lambda}\epsilon_{\lambda\rho\sigma\beta}F^{\rho\sigma}n^{\beta},\label{Sb5}
	\end{equation}
	i.e., it replays the form in Eq. (\ref{axion1}).

	If we use the $d$-dimensional
	replacements $k_{\alpha}k_{\beta}\to\frac{1}{d}\eta_{\alpha\beta}k^{2}$
	and $k_{\alpha}k_{\beta}k_{\gamma}k_{\delta}\to\frac{1}{d(d+2)}(\eta_{\alpha\beta}\eta_{\gamma\delta}+\eta_{\alpha\gamma}\eta_{\beta\delta}+\eta_{\alpha\delta}\eta_{\beta\gamma})$, with $d=4+\epsilon$,
	and afterward consider the $d\to4$ limit, the formally superficially
	divergent part of this contribution turns out to yield a non-zero
	finite result (we note that finiteness of superficially divergent
	contributions is a rather typical situation in LV theories, see, e.g., \cite{Colladay:1998fq,Gomes:2009ch}), which exactly cancels the finite
	contribution (\ref{Sb5}), resulting now in $\Sigma_{bn}=0$. Then,
	we conclude that this contribution is ambiguous. In analogy with calculations
	of the CFJ term, it is natural to expect that other regularizations
	could yield other results.
	
	There is one more way to prove the ambiguity of this contribution.
	We consider the following one-loop contribution to the effective action,
	involving the same two couplings present in Eq.\,\eqref{eq:contrib},
	\begin{equation}
		\Gamma=-i{\rm Tr}\ln(i\ds-m-e\As-\bs\gamma_{5}-\gamma_{5}\gamma_{\mu}\epsilon^{\mu\alpha\beta\nu}n_{\gamma}F_{\alpha\beta}),
	\end{equation}
	which can be rewritten as 
	\begin{equation}
		\Gamma=-i{\rm Tr}\ln(i\ds-m-e\As-\Bs\gamma_{5}),\label{eqsb5}
	\end{equation}
	where 
	\begin{equation}
		B^{\mu}=b^{\mu}-\epsilon^{\mu\alpha\beta\nu}n_{\gamma}F_{\alpha\beta}\thinspace.
	\end{equation}
	It is clear that the lower Lorentz violating and hence $B_{\mu}$-dependent
	contribution to this trace, apart from the non-axion contribution
	proportional to $(\epsilon^{\mu\alpha\beta\nu}n_{\gamma}F_{\alpha\beta})^{2}$,
	is proportional to $\epsilon^{\mu\nu\rho\sigma}B_{\mu}A_{\nu}\partial_{\rho}A_{\sigma}$.
	However, this expression will yield either the usual CFJ term without
	any impact of the nonminimal vertex, that is, the trivial scheme,
	or the contribution of third order in background fields which is also
	irrelevant for our purposes as we are interested in quadratic contributions
	only, in order to match\,\eqref{axion1}. That is to say, the possible
	relevant axion-photon contribution of first order in $A_{\mu}$ and
	in $B_{\mu}$ does not exist. Moreover, if the vector $B_{\mu}$ is not constant,
	the term $\epsilon^{\mu\nu\rho\sigma}B_{\mu}A_{\nu}\partial_{\rho}A_{\sigma}$
	is not gauge invariant. Therefore, the vanishing result (independent
	of the direction of vectors $b_{\mu}$ and $n_{\mu}$) is preferable
	in a certain sense, see the discussion in\,\cite{Altschul:2019eip}.
	
\begin{figure}[h]
		\centering{}\includegraphics[scale=0.9]{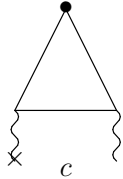} \caption{Contribution with a LV modification of one of legs.}
		\label{Fig3}
	\end{figure}

	The second appropriate dimension-5 operator in this case is $\bar{\psi}a_{F}^{(5)\mu\alpha\beta}\gamma_{\mu}F_{\alpha\beta}\psi$
	with $a_{F}^{(5)\mu\alpha\beta}=a^{\alpha}\eta^{\mu\beta}-a^{\beta}\eta^{\mu\alpha}$.
	The diagram responsible for generating the axion-photon contribution,
	in this case, with the $\bs\gamma_{5}$ insertion in the propagator,
	is represented in Fig.~\ref{Fig3}, which has been
	considered in~\cite{Borges:2013eda}.
	
	\vspace*{3mm}
	
	\vspace*{2mm}
	
	In this graph, the $\times$ sign means that the corresponding gauge
	field $A_{\nu}$ is replaced by $b^{\mu}F_{\mu\nu}$. We note that
	actually this contribution is described by the standard triangle graph
	that appears within studies of the CFJ term. Repeating the calculations
	that where already considered, for instance, in\,\cite{Borges:2013eda,Gomes:2007rv},
	we arrive at the following contribution 
	\begin{equation}
		\Sigma_{ab}=C_{0}\epsilon^{\alpha\beta\gamma\delta}b_{\alpha}a^{\mu}F_{\mu\beta}\partial_{\gamma}A_{\delta}=\frac{1}{2}C_{0}b_{\alpha}a^{\mu}\epsilon^{\alpha\beta\gamma\delta}F_{\mu\beta}F_{\gamma\delta},\label{axi2}
	\end{equation}
	with $C_{0}$ being a finite but arbitrary constant arising from the
	usual triangle diagram  contributing to the CFJ term, defined explicitly by the expression
		\begin{eqnarray}
			C_0\eta_{\alpha\beta}=2\int\frac{d^4k}{(2\pi)^4}\frac{\eta_{\alpha\beta}(k^2+3m^2)-4k_{\alpha}k_{\beta}}{(k^2-m^2)^2}
		\end{eqnarray}
		and, in various papers, found to be equal to $\frac{1}{4\pi^{2}}$, $\frac{3}{16\pi^{2}}$,
	zero, etc., within different regularization schemes (see \cite{Gomes:2007rv} and references therein for a detailed discussion of this  contribution).  
	We note that the same $C_{0}$ arose for the axion term in our previous
	paper \cite{Borges:2013eda}. It is clear that this term matches the
	axion term given by Eq.~(\ref{axion1}).
	
	So, besides the scenario presented in \cite{Borges:2013eda}, and
	the trivial scheme commented in section \ref{IS}, where one calculates
	the CFJ term considering $b_{\mu}=\partial_{\mu}\phi$, we found
	only three possibilities to obtain the axion term in the form of a
	finite one-loop correction. They are those generated by the $a_{F}^{(5)\mu\alpha\beta}\bar{\psi}\gamma_{\mu}F_{\alpha\beta}\psi$
	vertex together with the $\bs\gamma_{5}$ insertion in the propagator,
	resulting in Eq.\,(\ref{axi2}), by the $b_{F}^{(5)\mu\alpha\beta}\bar{\psi}\gamma_{5}\gamma_{\mu}F_{\alpha\beta}\psi$
	vertex together with the same $\bs\gamma_{5}$ insertion in the propagator,
	leading to Eq.\,(\ref{Sb5}), and by two nonminimal vertices $a^{\alpha}\bar{\psi}F_{\alpha\beta}\gamma^{\beta}\psi$
	and $b_{\alpha}\bar{\psi}\epsilon^{\alpha\beta\gamma\delta}F_{\beta\gamma}\gamma_{\delta}\psi$,
	resulting in Eq. (\ref{finamb}). We note that all these results are
	finite and ambiguous.
	
	\subsubsection{Second-order contributions with $\bs\gamma_{5}$ and $iD_{(\alpha}iD_{\beta)}$}
	
	Now, it remains to consider the possibilities for generating the axion
	based on terms proportional to $iD_{(\alpha}iD_{\beta)}$, with one
	$\bs\gamma_{5}$ insertion. We adopt the same reasoning as above.
	It is clear that contributions involving the quartic vertex from the
	term $\bar{\psi}iD_{(\alpha}iD_{\beta)}\psi$ (i.e., $\bar{\psi}A_{\alpha}A_{\beta}\psi$)
	cannot yield the axion form\,\eqref{axion1}, giving only derivative
	independent terms, which are further canceled with analogous terms
	arising from other Feynman diagrams. So, we take into account only
	triple vertices. Then, we consider graphs of the form given by Fig.~\ref{Fig1}(a)
	with one vertex now involving the factor $iD_{(\alpha}iD_{\beta)}$,
	while another vertex is minimal, and of the form given by Fig.~\ref{Fig1}(b)
	with a nonminimal LV insertion into the propagator, while both vertices
	are minimal, and the propagators are $b_{\mu}$-dependent and must
	be expanded in power series in $b_{\mu}$ up to the first order.
	
	By analogy with the previous discussions, it is easy to conclude that
	in both cases the axion-photon contributions will differ from zero only
	if odd-rank nonminimal Lorentz-breaking parameters are used. Moreover,
	the new vertex cannot involve $\gamma_{5}$ (otherwise, there will
	be no possibility to generate the Levi-Civita symbol since we already
	have the $\bs\gamma_{5}$ insertion). So, the only new vertex which
	could contribute to the axion term is the $a^{(5)\mu\alpha\beta}\bar{\psi}\gamma_{\mu}iD_{(\alpha}iD_{\beta)}\psi$.
	
	Thus, it remains to consider the last possibility, that is, to calculate
	contributions generated by $a^{(5)\mu\alpha\beta}\bar{\psi}\gamma_{\mu}iD_{(\alpha}iD_{\beta)}\psi$
	together with $\bs\gamma_{5}$, i.e., of first-order in $a^{(5)\mu\alpha\beta}$
	and $b_{\mu}$. So, effectively our Lagrangian looks like 
	\begin{equation}
		{\cal L}=\bar{\psi}(i\ds-m-\bs\gamma_{5}-e\As-a^{(5)\mu\alpha\beta}\bar{\psi}\gamma_{\mu}iD_{(\alpha}iD_{\beta)})\psi.
	\end{equation}
	
\begin{figure}[h]
		\centering{}\includegraphics[scale=0.9]{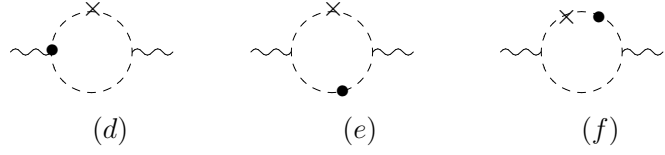} \caption{Contributions involving both a $a^{(5)\mu\alpha\beta}$ (represented
			by $\bullet$) and a $\bs\gamma_{5}$ (represented by $\times$) insertions.}
		\label{Fig4}
	\end{figure}

	In this case, we have the terms given by diagrams depicted in Fig.~\ref{Fig4}.
	
	\vspace*{3mm}
	
	\vspace*{3mm}
	
	Let us write the contributions arising from these graphs. For (d),
	we have 
	\begin{eqnarray}
		\Sigma_{ab1}(p) & = & ie^{2}a^{(5)\mu\alpha\beta}A_{\beta}(-p)A^{\lambda}(p)\int\frac{d^{4}k}{(2\pi)^{4}}\\
		& \times & {\rm tr}\Big[-p_{\alpha}\Big(\gamma_{\mu}(\ks+m)\bs\gamma_{5}(\ks+m)\gamma_{\lambda}(\ks+m)\ps(\ks+m)\nonumber \\
		& + & \gamma_{\mu}(\ks+m)\gamma_{\lambda}(\ks+m)\bs\gamma_{5}(\ks+m)\ps(\ks+m)\nonumber \\
		& + & \gamma_{\mu}(\ks+m)\gamma_{\lambda}(\ks+m)\ps(\ks+m)\bs\gamma_{5}(\ks+m)\Big)\frac{1}{(k^{2}-m^{2})^{4}}+\nonumber \\
		& + & 2k_{\alpha}\Big(\gamma_{\mu}(\ks+m)\bs\gamma_{5}(\ks+m)\gamma_{\lambda}(\ks+m)\ps(\ks+m)\ps(\ks+m)\nonumber \\
		& + & \gamma_{\mu}(\ks+m)\gamma_{\lambda}(\ks+m)\bs\gamma_{5}(\ks+m)\ps(\ks+m)\ps(\ks+m)\nonumber \\
		& + & \gamma_{\mu}(\ks+m)\gamma_{\lambda}(\ks+m)\ps(\ks+m)\bs\gamma_{5}(\ks+m)\ps(\ks+m)\nonumber \\
		& + & \gamma_{\mu}(\ks+m)\gamma_{\lambda}(\ks+m)\ps(\ks+m)\ps(\ks+m)\bs\gamma_{5}(\ks+m)\Big)\frac{1}{(k^{2}-m^{2})^{5}}\Big],\nonumber 
	\end{eqnarray}
	for (e), 
	\begin{eqnarray}
		\Sigma_{ab2}(p) & = & -ie^{2}A^{\mu}(-p)A^{\lambda}(p)a^{(5)\rho\alpha\beta}\int\frac{d^{4}k}{(2\pi)^{4}}k_{\alpha}k_{\beta}\nonumber \\
		& \times & {\rm tr}\Big[\gamma_{\mu}(\ks+m)\gamma_{\rho}(\ks+m)\gamma_{\lambda}\nonumber \\
		& \times & \Big((\ks+m)\ps(\ks+m)\bs\gamma_{5}(\ks-m)\ps(\ks+m)+(\ks+m)\bs\gamma_{5}(\ks+m)\ps(\ks+m)\ps(\ks+m)\nonumber \\
		& + & (\ks+m)\ps(\ks+m)\ps(\ks+m)\bs\gamma_{5}(\ks+m)\Big)\Big]\frac{1}{(k^{2}-m^{2})^{6}},
	\end{eqnarray}
	and finally, for (f), 
	\begin{eqnarray}
		\Sigma_{ab3}(p) & = & -ie^{2}A^{\mu}(-p)A^{\lambda}(p)a^{(5)\nu\alpha\beta}\int\frac{d^{4}k}{(2\pi)^{4}}k_{\alpha}k_{\beta}\\
		& \times & {\rm tr}\Big[\gamma_{\mu}(\ks+m)\bs\gamma_{5}(\ks+m)\gamma_{\nu}(\ks+m)\gamma_{\lambda}(\ks+m)\ps(\ks+m)\ps(\ks+m)\nonumber \\
		& + & \gamma_{\mu}(\ks+m)\gamma_{\nu}(\ks+m)\bs\gamma_{5}(\ks+m)\gamma_{\lambda}(\ks+m)\ps(\ks+m)\ps(\ks+m)\Big]\nonumber \\
		& \times & \frac{1}{(k^{2}-m^{2})^{6}}.\nonumber 
	\end{eqnarray}
	Taking into account that the axion-photon contribution has the form
	(\ref{axion1}), we require the $a^{(5)\mu\alpha\beta}$ to be completely
	characterized by a vector $a^{\mu}$. Since $a^{(5)\mu\alpha\beta}$
	is symmetric with respect to two last indices, we see that there are
	two possibilities to describe this tensor in terms of the unique vector
	needed to obtain (\ref{axion1}), that is, $a^{(5)\mu\alpha\beta}=a^{\mu}\eta^{\alpha\beta}$
	and $a^{(5)\mu\alpha\beta}=\eta^{\mu\alpha}a^{\beta}+\eta^{\mu\beta}a^{\alpha}$
	(in principle, one can introduce as well the completely symmetric
	case $a^{(5)\mu\alpha\beta}=\eta^{\mu\alpha}a^{\beta}+\eta^{\mu\beta}a^{\alpha}+\eta^{\alpha\beta}a^{\nu}$,
	but this is nothing more as the sum of the previous cases). As a consequence,
	it is natural to expect that the total result of these contributions
	is of the form 
	\begin{equation}
		\Sigma_{ab}=h_{1}\epsilon_{\mu\nu\lambda\rho}b^{\mu}F^{\nu\lambda}a_{\sigma}F^{\sigma\rho}+h_{2}\epsilon_{\mu\nu\lambda\rho}a^{\mu}F^{\nu\lambda}b_{\sigma}F^{\sigma\rho},\label{totres}
	\end{equation}
	with $h_1$ and $h_2$ being some numbers and $\Sigma_{ab}=\Sigma_{ab1}+\Sigma_{ab2}+\Sigma_{ab3}$. Then, straightforward computation allows to show that $\Sigma_{ab1}=0$, for an arbitrary form of $a^{(5)\mu\alpha\beta}$.  Further, the explicit calculations
	of integrals for $\Sigma_{ab2}$ and $\Sigma_{ab3}$ show that both
	if $a^{(5)\mu\alpha\beta}=\eta^{\mu\alpha}a^{\beta}+\eta^{\mu\beta}a^{\alpha}$
	and if $a^{(5)\mu\alpha\beta}=a^{\mu}\eta^{\alpha\beta}$, the axion-photon
	contributions diverge and are equal for (e) and (f) diagrams. So,
	we can write down the final result for $\Sigma_{ab}$, the same for
	$a^{(5)\mu\alpha\beta}=\eta^{\mu\alpha}a^{\beta}+\eta^{\mu\beta}a^{\alpha}$
	and $a^{(5)\mu\alpha\beta}=a^{\mu}\eta^{\alpha\beta}$, as follows:
	\begin{equation}
		\Sigma_{ab}=\frac{e^{2}}{6\pi^{2}\epsilon}\epsilon^{\mu\nu\lambda\rho}(-2b_{\mu}a^{\alpha}+a_{\mu}b^{\alpha})F_{\alpha\lambda}F_{\rho\nu}\thinspace,\label{bDD}
	\end{equation}
	with $\epsilon=d-4$. We see that this result is less advantageous in comparison with the
	previous schemes as it requires to introduce the axion term from the
	very beginning, in order to achieve multiplicative renormalizability. 
	
	To close this section, let us briefly discuss the problem of ambiguities in our theory. As it is known, see, e.g., \cite{BaetaScarpelli:2013rmt,BaetaScarpelli:2015ykr} and references therein, ambiguities arise naturally when a superficially divergent integral turns out to be finite, and, in certain cases, they can signalize the presence of anomalies. For example, the ambiguity of the CFJ term is related with the Adler-Bell-Jackiw (ABJ) anomaly, and the ambiguity of the four-dimensional gravitational Chern-Simons term is related with gravitational anomalies. Therefore, the ambiguity of the axion term arising within several schemes of its calculation can be treated as an argument in favor of the existence of some new anomalies related with the axion. From another side, the ABJ anomaly equal to $\tilde{F}^{\mu\nu}F_{\mu\nu}$ multiplied by a some number yields the same form proportional $\vec{E}\cdot\vec{B}$ as the axion-photon term $b^{\alpha}\tilde{F}_{\alpha\mu}a_{\beta}F^{\beta\mu}$. Therefore, it is natural to treat the ambiguity of the axion term as a consequence of the ABJ anomaly.
	
	%\section{Phenomenological estimations}
	\section{Constraints on the Lorentz violation parameters from the axion-photon coupling}  \label{lvplimits}

 The limits from direct searches, astrophysics and cosmology for axions/ALPs on the axion-photon 
 coupling allow us to constrain the related LV parameters. In a more general context, the total effective coupling would still receive contributions out of LI operators so that $g_{\phi\gamma}= g_{\phi\gamma}^{(LI)} + g_{\phi\gamma}^{(LV)}$. An upper bound for the LV can be obtained assuming that $g_{\phi\gamma}\approx g_{\phi\gamma}^{(LV)}$. The limits over the axion-photon 
 coupling dependent on axion mass~\cite{DiLuzio:2020wdo,ParticleDataGroup:2022pth}. For example, the CAST experiment puts the limit $|g_{\phi\gamma}| < 6.6\times10^{-11}$ GeV$^{-1}$ for axion mass $m_\phi<0.02$ eV~\cite{CAST:2017uph}. Taking into account the coupling from Eq. (\ref{axion1}), generated by the LV operators  $d^{\nu}F_{\mu\nu}\overline{\psi}\gamma^{\mu}\psi$ and  $b_{\mu}\phi\bar{\psi}\gamma_{5}\gamma^{\mu}\psi$, one has  $g_{\phi\gamma}=4\sqrt{\pi\alpha}C(b\cdot d)\approx 0.6\, C(b\cdot d)$, in which $\alpha=1/137$. Thus, the CAST limits implies that $|C(b\cdot d)| < 1.1\times10^{-10}$ GeV$^{-1}$. 

In the case of the axion-photon effective Lagrangian generated by LV operators $\bar{\psi}a^{\alpha}\phi F_{\alpha\beta}\gamma^{\beta}\psi$ and $\bar{\psi}\epsilon^{\alpha\beta\gamma\delta}b_{\alpha}F_{\beta\gamma}\gamma_{\delta}\psi$, the coupling that follows from Eq. (\ref{finamb}) is $g_{\phi\gamma}=2C_1 m^2(a\cdot b)$, with $a^{\alpha}$ and $b_{\alpha}$ having dimension of inverse of mass squared and inverse of mass, respectively. Then, the constraint from CAST turns out to be $|C_1 m^2(a\cdot b)| < 3.3\times10^{-11}$ GeV$^{-1}$. As we see, the higher the fermion mass $m$, the stronger is the  constraint over the product $|C_1 (a\cdot b)|$ of the LV parameters. As an example, for a fermion of mass $m=1$ TeV,  the limit would be $|C_1 (a\cdot b)| < 3.3\times10^{-17}$ GeV$^{-3}$.  While, in Ref.~\cite{Kostelecky:2008ts} the operator $\phi \tilde{F}F$ is not considered, we note that, in principle, the estimations we have done can serve for obtaining better results for constraining non-minimal LV parameters. Finally, the proportionality of $g_{\phi\gamma}^{(LV)}$ with $m^2$ seems to pose a problem for this coupling constant once the it does not lead to a decoupling as $m$ gets large, contrary to the typical $g_{\phi\gamma}^{(LI)}$ obtained through integration of heavy fermions in a renormalizable LI theory. But, we note that,  actually, the theory in Eq.~(\ref{twononmin}) is an effective one, valid below an energy scale $\Lambda_{LV}$ related to the LV parameters $a^{\alpha}\,(\sim 1/\Lambda_{LV}^2)$ and $b_{\alpha}\,(\sim 1/\Lambda_{LV})$ out of the non-renormalizable operators  (the constant $C_1$ does not play here an essential role being of the order of the unit). In this vein, the fermion mass is assumed such that $m\ll \Lambda_{LV}$ in order to have a consistent low energy effective Lagrangian describing the axion-photon interaction.

	\section{Summary\label{SUM}}
	
	We have demonstrated explicitly that the axion-photon interaction term arises within certain dimension-5 Lorentz-breaking extensions as a quantum correction. The most interesting situations are those where the axion-photon term  is finite, so one does not need to worry about its renormalization, hence, introducing it from the beginning is unnecessary.
	
	However, as we have noted, besides the trivial scheme where the axion contribution is obtained from the CFJ term and the scheme described in \cite{Borges:2013eda} and presented here in Fig.~\ref{Fig3} (see Eq.~(\ref{axi2})), there are only a few possibilities for the radiative generation of the axion-photon interaction term. These take into account: first, two nonminimal vertices involving axial and usual vectors (see Eq.~(\ref{finamb})); second, one minimal vertex $\bs\gamma_{5}$ insertion and one vertex involving $\bar{\psi}b_{F}^{(5)\mu\alpha\beta}\gamma_{5}\gamma_{\mu}F_{\alpha\beta}\psi$ (see Eq.~(\ref{Sb5})); third, a $\bs\gamma_{5}$ insertion and the $a^{(5)\mu\alpha\beta}\bar{\psi}\gamma_{\mu}iD_{(\alpha}iD_{\beta)}\psi$ vertex, with this last case representing a divergent contribution (see Eq.~(\ref{bDD})).
	
	One can understand this scarcity in the possibilities in the following terms: the axion-photon contributions are always superficially divergent, as follows from the degree of divergence of the corresponding Feynman diagrams, therefore, they can be finite only if they are generated by an ambiguous Feynman diagram, and there were only two such one-loop diagrams known up to now, the triangle one studied in \cite{Borges:2013eda}, and that one studied in \cite{Gomes:2007rv}. Besides these diagrams, we found that the lower contribution in Eq. (\ref{Sb5}) arising from the one-loop effective action in Eq. (\ref{eqsb5}) is also ambiguous. In all other cases, the axion-photon terms either diverge or are equal to zero. It must be also noted that in many cases the axion-photon terms, besides displaying divergences, are of the second order in the Lorentz violating parameters, thus expected to be highly suppressed.
	
	Now, we formulate the prescriptions for possible schemes allowing us to generate axion-photon interactions terms for generic higher-dimension operators. We can consider two situations: (i) contributions of the first order in Lorentz-breaking parameters; (ii) contributions of the second order in Lorentz-breaking parameters. However, as we already argued, the first case is trivial, effectively replaying the calculation of the usual CFJ term. The second case requires attention. Then, we note that the axion-photon terms actually can be generated from schemes where the LV vertices involve either one Levi-Civita symbol or one $\gamma_{5}$ matrix, but not together. Also, we can consider an odd-rank Lorentz-breaking tensor to obtain a CFJ-like term that is transformed to the axion-photon term after integration by parts, and an even-rank tensor to obtain an explicit axion-photon term involving two derivatives. It should be noted nevertheless that it seems to be very improbable to obtain finite axion-photon  terms for higher-dimension operators since finite results can arise from non-renormalizable theories only in very rare situations. Nevertheless, such corrections were shown to arise due to the cancellation of divergences like those that occurred in \cite{Gomes:2009ch}. We have demonstrated that within all schemes considered in this work, the non-zero axion-photon terms either diverge, which is a less valuable result since it requires introducing the axion term at the tree level to achieve multiplicative renormalizability, or are ambiguous. This last case could signalize some profound relation between axions and anomalies.
	
	The natural continuation of this study would be an analysis of the possibilities for generating axion-photon interaction term for operators with dimensions six  and, perhaps, from higher loops. However, these calculations will be much more involved, and the absence of CFJ contributions in higher loops argued in \cite{Brito:2020nvb} will imply the absence of higher-loop axion contributions, at least in certain cases. We plan to perform a more detailed discussion of higher-order contributions to the axion-photon interaction term, as well as of other contributions generated by higher-dimension LV terms in our next papers.
	
	\bigskip{}
	
	\textbf{Acknowledgments}: The authors are grateful to J. C. Rocha
	for important discussions. This study was financed in part by Conselho
	Nacional de Desenvolvimento Científico e Tecnológico (CNPq), via the
	grants 305802/2019-4 (A.G.D), 305967/2020-7 (A.F.F) and 301562/2019-9
	(A.Yu.P.).

	\bibliography{myrefs}
	
\end{document}